\documentclass{PoS}

\def\LP{\left(}		% left parenthesis
\def\RP{\right)}	% right parenthesis
\def\PAR#1#2{ {{\partial #1}\over{\partial #2}} }

\def\BNE{\begin{equation}}
\def\ENE{\end{equation}}
\def\BE{\begin{displaymath}}
\def\EE{\end{displaymath}}
\def\BEA{\begin{eqnarray*}}
\def\EEA{\end{eqnarray*}}
\def\BNEA{\begin{eqnarray}}
\def\ENEA{\end{eqnarray}}

\newcommand{\R}[1]{{#1}}

\newcommand{\B}[1]{{#1}}

\title{Pseudoscalar meson physics with four dynamical quarks}

\ShortTitle{Pseudoscalar meson physics with four dynamical quarks}

\author{
A.~Bazavov,$^a$
C.~Bernard,$^b$
C.~Bouchard,$^c$
C.~DeTar,$^d$
D.~Du,$^e$
A.X.~El-Khadra,$^e$
J.~Foley,$^d$
E.D.~Freeland,$^f$
E.~Gamiz,$^g$
Steven~Gottlieb,$^h$
U.M.~Heller,$^i$
J.E.~Hetrick,$^j$
J.~Kim,$^k$
A.S.~Kronfeld,$^l$
J.~Laiho,$^m$
L.~Levkova,$^d$
M.~Lightman,$^b$
P.B.~Mackenzie,$^l$
E.T.~Neil,$^l$
M.~Oktay,$^d$
J.N.~Simone,$^l$
R.L.~Sugar,$^n$
\speaker{D.~Toussaint,}\nolinebreak $^k$
R.S.~Van~de~Water,$^{a,l}$
and
R.~Zhou$^h$
[Fermilab Lattice and MILC Collaborations]
\\
\llap{$^a$} Department of Physics, Brookhaven National Laboratory\thanks{Operated by Brookhaven Science Associates, LLC, under
Contract No.~DE-AC02-98CH10886 with
the U.S. Department of Energy.}, Upton, NY 11973, USA\\
\llap{$^b$} Department of Physics, Washington University, St. Louis, MO 63130, USA\\
\llap{$^c$} Department of Physics, The Ohio State University, Columbus, OH 43210, USA\\
\llap{$^d$} Physics Department, University of Utah, Salt Lake City, UT 84112, USA\\
\llap{$^e$} Physics Department, University of Illinois, Urbana,  IL 61801, USA\\
\llap{$^f$} Department of Physics, Benedictine University, Lisle, IL 60532, USA\\
\llap{$^g$} CAFPE and Departamento de Fisica Te\,orica y del Cosmos, Universidad de Granada, Granda, Spain\\
\llap{$^h$} Department of Physics, Indiana University, Bloomington, IN 47405, USA\\
\llap{$^i$} American Physical Society, One Research Road, Ridge, NY 11961, USA\\
\llap{$^j$} Physics Department, University of the Pacific, Stockton, CA 95211, USA\\
\llap{$^k$} Physics Department, University of Arizona, Tucson, AZ 85721, USA\\
\llap{$^l$} Fermi National Accelerator Laboratory\thanks{Operated by Fermi Research Alliance, LLC, under Contract
No.~DE-AC02-07CH11359 with
the U.S. Department of Energy.}, Batavia, IL 60510, USA\\
\llap{$^m$} SUPA, School of Physics and Astronomy, University of Glasgow, Glasgow G12 8QQ, UK\\
\llap{$^n$} Department of Physics, University of California, Santa Barbara, CA 93106, USA\\
E-mail: 
\email{doug@physics.arizona.edu},
}

\abstract{
We present preliminary results for light, strange and charmed pseudoscalar meson
physics from simulations using four flavors of dynamical quarks with the highly
improved staggered quark (HISQ) action.  These simulations include lattice spacings
ranging from 0.15 to 0.06 fm, and sea-quark masses both above and at their
physical value.  The major results are
charm meson decay constants $f_D$,  $f_{D_s}$ and $f_{D_s}/f_D$ and
ratios of quark masses.
This talk will focus
on our procedures for finding the decay constants on each ensemble, the continuum extrapolation, and
estimates of systematic error.
          }

\FullConference{The 30th International Symposium on Lattice Field Theory - Lattice 2012\\
		June 24-29, 2012\\
		Cairns, Australia}

\begin{document}

\section{Introduction} \vspace{-2.0mm}

Lattice calculations of pseudoscalar meson decay constants, when combined
with measured nonleptonic decay rates, can be used to extract CKM matrix
elements and test the Standard Model.
The lattice calculation includes a determination of the quark masses ---
fundamental parameters of the Standard Model.
Here we present preliminary results for the charm meson decay constants $f_D$ and
$f_{D_s}$, and the ratios of quark masses $m_c/m_s$ and $m_u/m_d$.

These calculations use lattices generated with the one-loop and tadpole improved
Symanzik gauge action~\cite{SYMANZIK} and the HISQ fermion action~\cite{HISQACTION}.
We include four flavors of dynamical sea quarks:  degenerate up and down,
strange, and charm.
Iterated smearing in the HISQ action reduces taste violations by a factor
of three relative to the asqtad action.  An improved charm-quark dispersion
relation allows us to treat it the same way as the lighter quarks.
Our lattice ensembles include ensembles with approximately physical light-quark
masses, with volumes as large as $5.6^3$ fm$\null^3$.
For more details, see Refs~\cite{SCALING09,HISQLATTICE10,ENSEMBLES12}.
The parameters of the ensembles used in this calculation are tabulated in
J.~Kim's talk~\cite{JJTALK}.
In the first stage of analysis we fit pseudoscalar meson two-point
correlators to extract masses and amplitudes from random wall operators
for a set of valence-quark masses. (See Ref.~\cite{JJTALK}.)
In the second stage of the analysis, 
%described in section \ref{STAGE2},
we interpolate or extrapolate in valence-quark
masses to find the tuned valence masses on each ensemble, and the decay
constants evaluated at these valence masses, but at the sea-quark masses
and lattice spacing of the individual ensemble.
Then we combine the results of the different ensembles to make a continuum extrapolation
and corrections for mistuned sea-quark masses.  
%This stage is described in section \ref{STAGE3}.

We use the pion decay constant $f_\pi$ to set the
scale.  Since this is determined from the same set of correlators as the
charm meson decay constants, the analysis is self-contained in the sense that
we do not need an intermediate quantity such as $r_0$ or $r_1$ found
in a separate analysis.  This reduces the error from determining
the lattice spacing, and simplifies the error analysis.
Ensembles with physical light-quark masses dominate the analysis and
allow a simple interpolation to adjust for light-quark mass corrections.
However, it is likely that use of
staggered chiral perturbation theory will lead to better controlled fits,
and J. Komijani's talk describes progress in this direction~\cite{KOMIJANITALK}.

\section{Decay constants on each ensemble} \vspace{-2.0mm}
\label{STAGE2}

%\begin{figure}
%\vspace{-0.5in}
%\begin{center}
%\begin{tabular}{ll}
%\hspace{-0.3in}\includegraphics[width=0.52\textwidth]{fdds_vs_a_fpi_1023.pdf} &
%\hspace{-0.0in}\includegraphics[width=0.52\textwidth]{fdds_vs_a_fpi_fits.pdf} \\
%\end{tabular}
%\end{center}
%\vspace{-1.2in}
%\caption{
%\label{fig:fdds_fits_fpi}
%$f_D$ and $f_{D_s}$ on the different ensembles.  in the right panel, fits that are
%quadratic and linear in $a^2$ are superimposed on the data.  More details are given
%in the text.
%\vspace{-0.15in}
%}
%\end{figure}

\begin{figure}
\vspace{-0.5in}
\begin{center}
\begin{tabular}{l}
\hspace{-0.0in}\includegraphics[width=0.52\textwidth]{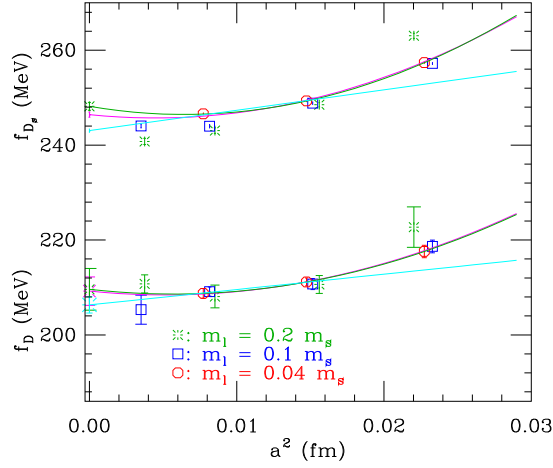} \\
\end{tabular}
\end{center}
\vspace{-1.2in}
\caption{
\label{fig:fdds_fits_fpi}
$f_D$ and $f_{D_s}$ on the different ensembles.  Fits that are
quadratic and linear in $a^2$, or use only physical quark mass ensembles are superimposed on the data.  More details are given
in the text.
\vspace{-0.15in}
}
\end{figure}

In the first stage of the analysis~\cite{JJTALK} we determine the masses and amplitudes
for the two-point pseudoscalar correlators
for a set of valence quark masses on each ensemble.  These masses include
$0.9$ and $1.0$ times the sea charm quark mass, $0.8$ and $1.0$ times the sea strange-quark mass,
and a range of lighter masses.  Since the sea-quark masses were estimated before
the production runs began, they are inevitably slightly mistuned.  In the
second stage of the analysis we determine tuned quark masses for each ensemble, and
evaluate the decay constants at these valence-quark masses.  Our primary analysis
uses $f_\pi$ to fix the lattice spacing, and proceeds as follows.  To estimate statistical
errors, this whole procedure is done inside a jackknife resampling.

\begin{enumerate}
\setlength{\topsep}{0mm}
\setlength{\parsep}{0mm}
\setlength{\itemsep}{-1.8mm}
\vspace{-3.0mm}
%\item{\B{Notation:} $m_A$, $m_B$ = valence masses, $m_s$, $m_l$, $m_c$ = tuned valence masses.}
\item{Interpolate or extrapolate in the light valence-quark mass $m_l$ to the point where
    ${M_\pi}/{f_\pi}$ has its physical\footnote{adjusted for E\&M and finite size --- see later}
    value.  This fixes $am_l$ and the lattice spacing $a$.}
\item{Interpolate or extrapolate in valence strange-quark mass $am_s$ to where 
    $2M_K^2-M_\pi^2$ has its 
\vspace{-0.05mm}    physical value.  This fixes $m_s$.}
\item{Interpolate or extrapolate to the charm valence mass where $M_{D_s}$ is correct.  This fixes $m_c$.}
\item{Find $m_d-m_u$ from E\&M adjusted $K^0-K^+$ mass difference~\cite{EMREFERENCES} }
\item{Find, by interpolation or extrapolation, $f_K$ at the adjusted up-quark mass.}
\item{Find $f_D$ and $M_D$ (a check) at the adjusted down and charm masses.}
\item{Find $f_{D_s}$ at the adjusted strange and charm masses.}
\item{Find $M_{\eta_c}$ (check) at the adjusted charm-quark mass.}
\end{enumerate}
\vspace{-3.0mm}

%An alternate tuning procedure, used to estimate systematic errors, uses $f_K$ as the
%length standard, beginning with:
%
%\begin{enumerate} 
%\setlength{\topsep}{0mm}
%\setlength{\parsep}{0mm}
%\setlength{\itemsep}{-1mm}
%
%\item{``FK tuning'': Using $m_l$ equal to the light sea-quark mass, interpolate/extrapolate in $m_s$ such
%that
%    $\frac{2M_K^2-M_\pi^2}{f_K^2}$ has its
%    physical%\footnote{\small adjusted for E\&M and finite size --- see later}
%    value.  This fixes $am_s$ and $a$}
%\item{Find $am_l$ where $M_\pi^2$ has its physical value.  This fixes $m_l$.}
%\end{enumerate}

We have also used a tuning procedure where we replaced $f_\pi$ and $M_\pi$ by
the decay constant, $f_{p4s}$, and mass, $M_{p4s}$, of a fictitious meson with degenerate valence
quark masses at 0.4 times the strange-quark mass, using the values for
$f_{p4s}$ and $M_{p4s}$ in MeV determined from three flavor asqtad
data.  Finally, we varied the procedure by first fixing the lattice
spacing from the static-quark potential, and then tuning the valence
quark masses as described above.

In Eq.~\ref{eqn:bestensemble} we list results from the above procedure for the most influential ensemble ($a \approx 0.09$ fm with
physical masses).  These results are adjusted for finite size and and
some electromagnetic effects, but are not adjusted for mistuned sea-quark
masses or extrapolated to the continuum limit.
The errors here are statistical only but, as discussed above, errors from scale setting
and valence-quark mass tuning are included in the statistical error here.
The $D^+$, $D^0$ and $\eta_c$ masses are not used in the tuning, so they can be
compared to their experimental values, shown in parentheses.

%\hspace{-0.2in}
%\begin{tabular}{lll}
%$a = 0.08792(10)$ fm & & \\
%$am_l = 0.001333(5)$ & $am_s=0.03648(11)$ & $am_c=0.4323(7)$ \\
%%$m_u/m_d = 0.480(6)$\footnote{sensitive to E\&M adjustments!!} & $m_s/m_l = 27.36(3)$ & $m_c/m_s=11.851(17)$  \\
%$m_u/m_d = 0.480(6)^*$ & $m_s/m_l = 27.36(3)$ & $m_c/m_s=11.851(17)$  \\
%\multicolumn{2}{l}{$f_K=155.12(22)$ MeV (\B{cf 156.1})} & \\
%%\multicolumn{3}{l}{$M_{D_0}=1867.8(1.5)$ MeV (\B{cf 1864.8})}  \\
%%\multicolumn{3}{l}{$M_{D^+}=1870.2(1.2)$ MeV (\B{cf 1869.6})}  \\
%\multicolumn{2}{l}{$M_{D_0}=1867.8(1.5)$ MeV (\B{cf 1864.8})}  &
%\multicolumn{1}{l}{$M_{D^+}=1870.2(1.2)$ MeV (\B{cf 1869.6})}  \\
%\multicolumn{2}{l}{$M_{\eta_c}=2980.2(3)$ MeV (\B{cf 2980.3(1.2)})} & \\
%\R{$f_D=208.75(1.03)$ MeV} & \R{$f_{D_s}=246.60(20)$ MeV} & \R{$f_{D_s}/f_D=1.181(6)$} \\
%\end{tabular}

\vspace{2mm}
\hspace{-0.15in}
\begin{tabular}{llll}
$a = 0.08792(10)$ fm & $am_l = 0.001333(5)$ & $am_s=0.03648(11)$ & $am_c=0.4323(7)$ \vspace{+1pt}\\
%$m_u/m_d = 0.480(6)$\footnote{sensitive to E\&M adjustments!!} & $m_s/m_l = 27.36(3)$ & $m_c/m_s=11.851(17)$  \vspace{+1pt}\\
$m_u/m_d = 0.480(6)$ & $m_s/m_l = 27.36(3)$ & $m_c/m_s=11.851(17)$  & \vspace{+1pt}\\
\multicolumn{4}{l}{$f_K=155.12(22)$ MeV }\vspace{+1pt}\\
%\multicolumn{3}{l}{$M_{D_0}=1867.8(1.5)$ MeV (\B{cf 1864.8})}  \vspace{+1pt}\\
%\multicolumn{3}{l}{$M_{D^+}=1870.2(1.2)$ MeV (\B{cf 1869.6})}  \vspace{+1pt}\\
\multicolumn{2}{l}{$M_{D_0}=1867.8(1.5)$ MeV (\B{cf 1864.8})}  &
\multicolumn{2}{l}{$M_{D^+}=1870.2(1.2)$ MeV (\B{cf 1869.6})}  \vspace{+1pt}\\
\multicolumn{2}{l}{$M_{\eta_c}=2980.2(3)$ MeV (\B{cf 2980.3(1.2)})} & & \vspace{+1pt}\\
\multicolumn{4}{l}{\R{$f_D=208.75(1.03)$ MeV} \hspace{0.5in} \R{$f_{D_s}=246.60(20)$ MeV} \hspace{0.5in} \R{$f_{D_s}/f_D=1.181(6)$}
}\vspace{+1pt}\\
\end{tabular}\\
%$*$ sensitive to E\&M adjustments, see later!
\vspace{-5mm}\BNE \label{eqn:bestensemble} \null \ENE 

\section{Continuum extrapolation} \vspace{-2.0mm}
\label{STAGE3}

The results for $f_D$ and $f_{D_s}$ on each ensemble are shown in
Fig.~\ref{fig:fdds_fits_fpi}.  In this figure the physical quark mass ensembles,
labelled as ``$m_l=0.04\,m_s$'', have the smallest statistical errors.  This is due
to their larger physical volumes,
%OMIT REST OF THIS SENTENCE?
where the random wall and
Coulomb wall sources, together with the average over spatial position of the sink
operator, result in better statistics.
Errors are
large on some of the $m_l=0.2\,m_s$ ensembles because the lightest valence-quark
mass used on these ensembles was $0.1\,m_s$, so a significant extrapolation in valence
light-quark mass was necessary.  When these points are fit to functions of lattice spacing
and sea-quark mass, the physical quark mass points dominate the fit.  Indeed, we could
get reasonable results by simply using the physical quark mass ensembles.  In practice,
we use the physical quark mass and the $m_l=0.1\,m_s$ ensembles, where the
$m_l=0.1\,m_s$ ensembles allow correction for mistuning of the light sea-quark masses in the physical
quark mass ensembles and, because of the $m_l=0.1\,m_s$ ensemble
at lattice spacing $a\approx 0.06$ fm, help to determine the dependence on the lattice
spacing.

Figure~\ref{fig:fdds_fits_fpi} shows three different fits to this
data.  The magenta line uses all four of the lattice spacings, and is quadratic
in $a^2$ and linear in light sea-quark mass.
% \BNE f_D(a,m_{l,\mathrm{sea}}) = f_D(a=0,m=m_{phys}) + C_1 a^2 + C_2 (a^2)^2 + C_m ( m_{l,\mathrm{sea}}/m_{phys} - 1 ) \ENE
% The form of the mass term is chosen so that the fit at $C_m=0$, plotted in
% Fig.~\ref{fig:fdds_fits_fpi}, is the fit form at the physical light-quark mass.
The straight cyan line in Fig.~\ref{fig:fdds_fits_fpi} omits the $a=0.15$ fm points, and
is linear in both $a^2$ and $m_{l,\mathrm{sea}}$.
The green line is a quadratic fit using only the physical quark mass points, with
small adjustments to correct for sea-quark mass mistuning.
The symbols at $a=0$ are the continuum limits of these fits, showing the statistical error.
The size of these symbols is proportional to the $p$-value of the fit, with the symbol size
in the graph legend corresponding to 50\%.  
Since we expect both order $a^2$ and $a^4$ corrections to the data, and since some other
accurately determined quantities such as the mass of the $\eta_c$ clearly need $a^4$
terms to fit the results, we take the quadratic fit results as our central value,
with the larger of the difference between the quadratic and linear fits or the difference
between the quadratic and physical-quark-mass-only fits as an estimate of the
systematic error coming from our choice of fit form.

\begin{figure}
\vspace{-0.5in}
\begin{center}
\begin{tabular}{ll}
\hspace{-0.3in}\includegraphics[width=0.52\textwidth]{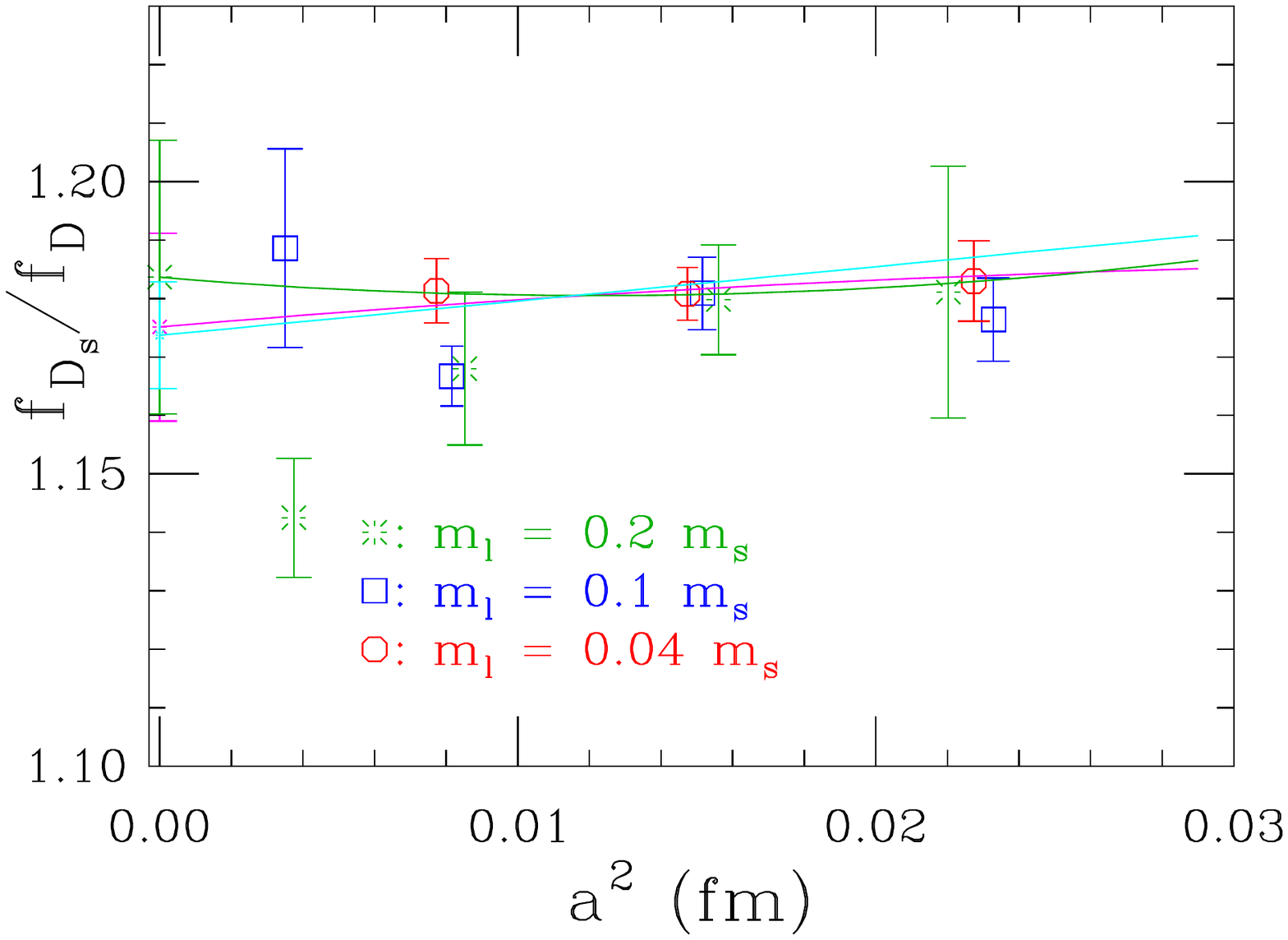} &
\hspace{-0.0in}\includegraphics[width=0.52\textwidth]{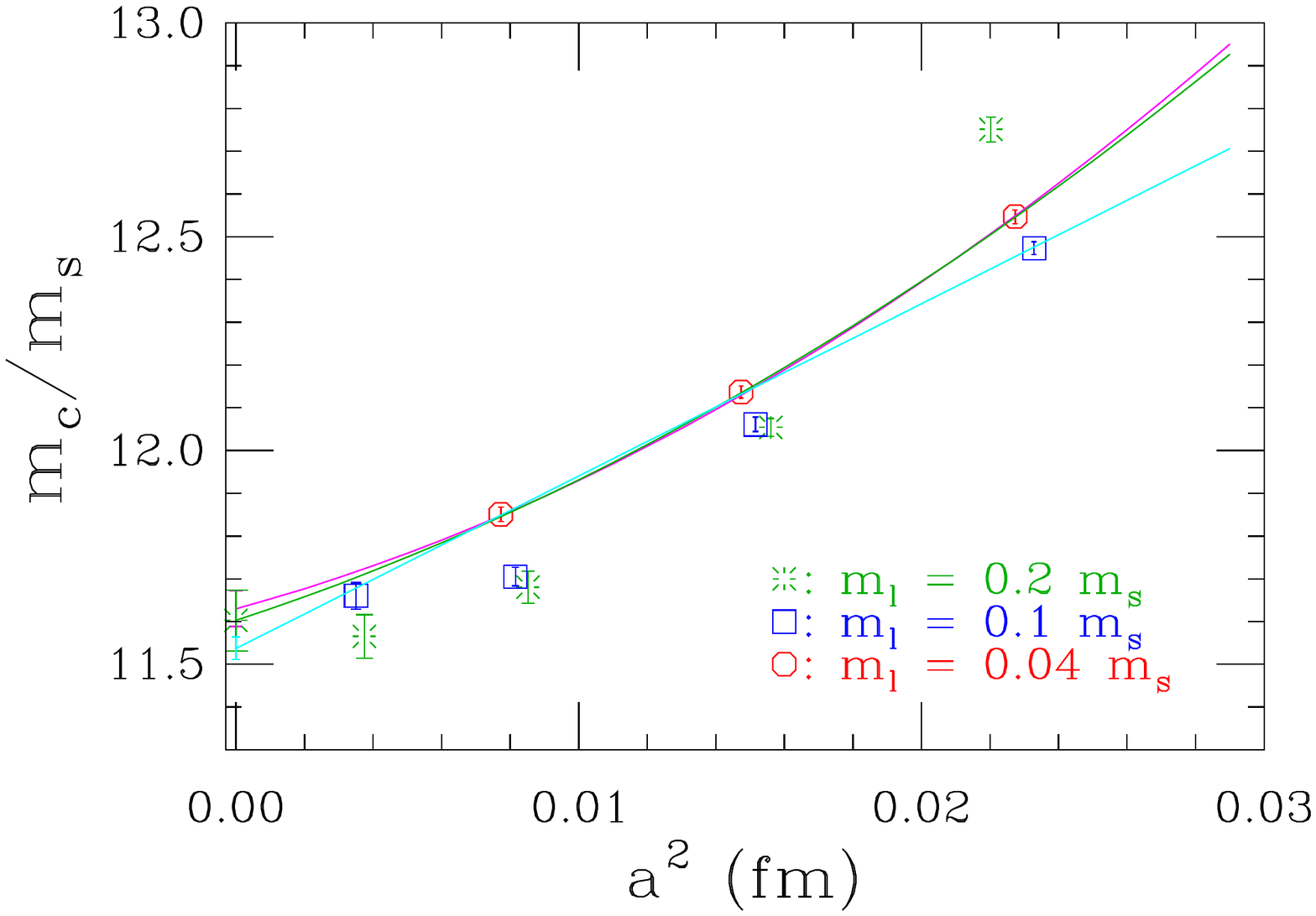} \\
\end{tabular}
\end{center}
\vspace{-1.6in}
\caption{
The ratios $f_{D_s}/f_D$ (left panel) and $m_c/m_s$ (right panel) on the different
ensembles.  The superimposed fit functions are the same forms as in Fig.~\protect\ref{fig:fdds_fits_fpi}.
\label{fig:tworatios_vs_a_fpi}
\vspace{-0.15in}
}
\end{figure}

Figure~\ref{fig:tworatios_vs_a_fpi} shows, with the same notation, the ratios $f_{D_s}/f_D$
and $m_c/m_s$ on each ensemble, together with fits to the same functional forms and
ranges of data as in Fig.~\ref{fig:fdds_fits_fpi}.

% try side-by-side figures

\begin{figure}
\vspace{-0.5in}
\begin{center}
\begin{tabular}{ll}
\hspace{-0.3in}\includegraphics[width=0.55\textwidth]{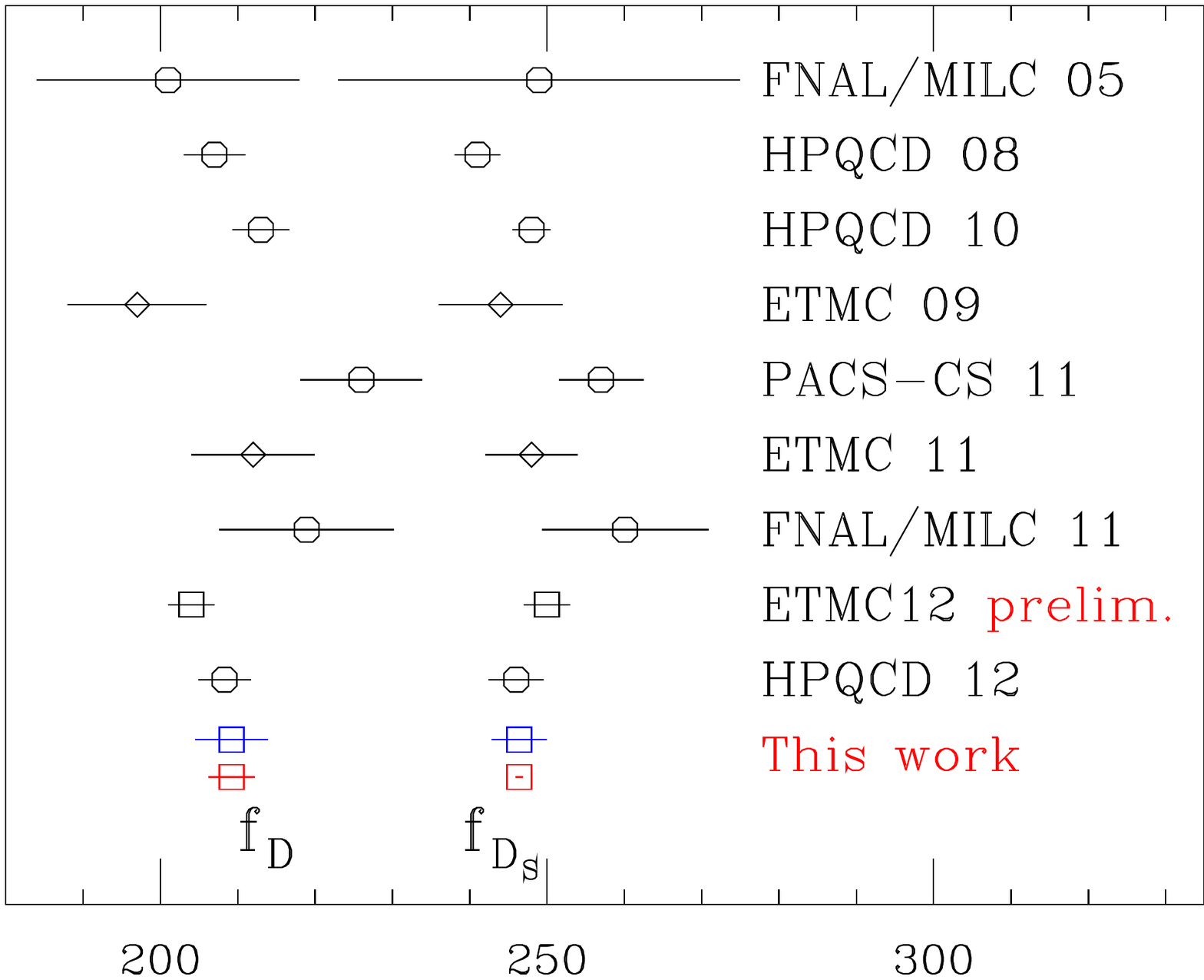} &
\hspace{-0.3in}\includegraphics[width=0.55\textwidth]{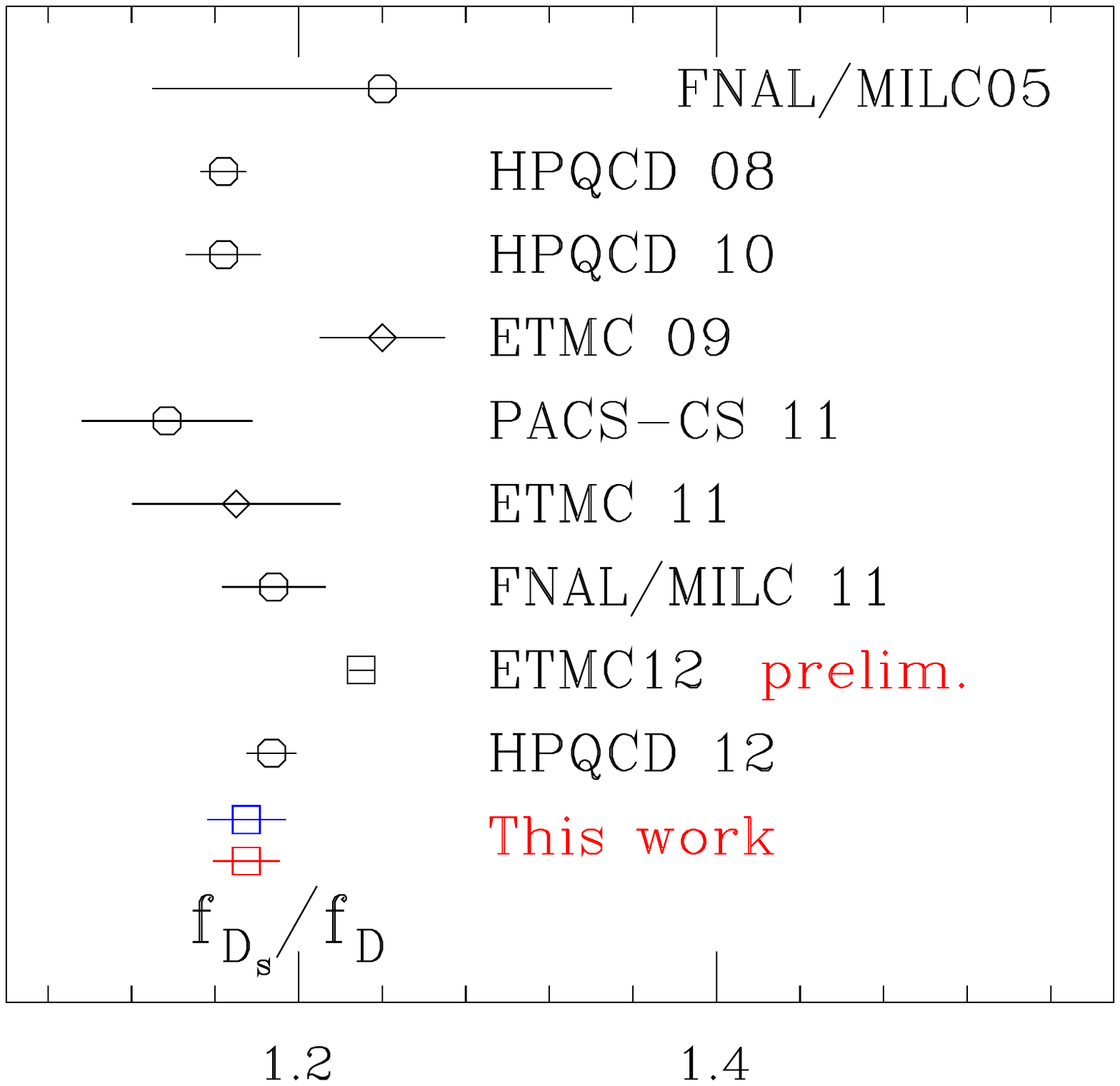} \\
\end{tabular}
\end{center}
\vspace{-1.3in}
\caption{
Comparison of lattice results for $f_D$ and $f_{D_s}$ (left panel) and for $f_{D_s}/f_D$
(right panel). 
%Results are from Refs.~\protect\cite{FNAL95,HPQCD08,HPQCD10,ETMC09,
%PACS-CS11,ETMC11,FNAL11,ETMC12,HPQCD12} and this work.
Results are from Ref.~\protect\cite{ALLFD}
and this work.
Diamonds represent results with 2, octagons with 2+1, and
squares with 2+1+1 dynamical flavors.
For this work the red error bars (lower) are statistical only, while
the blue error bars (higher) include systematic error estimates.
\label{fig:othervalues}
\vspace{-0.15in}
}
\end{figure}

\section{Systematic errors} \vspace{-2.0mm}
\label{SYSTEMATIC}

Statistical errors from the jackknife analysis incorporate errors in
lattice spacing determination and valence-quark mass tuning,
since this tuning is redone for each jackknife sample.
%Remaining systematic effects in our error budget include: excited state contamination, finite
%size effects, electromagnetic
%effects, effects of choices made in our scale setting and mass tuning
%procedures, and effects of choices made in our fits for the continuum
%extrapolation and sea-quark mass mistuning.
The following additional systematic errors are included in our error budget.

Effects of excited states in the two-point correlators were estimated by
varying the time range in the fits and the priors for excited state masses
with reasonable ranges~\cite{JJTALK}.

\begin{figure}
\vspace{-0.5in}
\begin{center}
\begin{tabular}{ll}
\hspace{-0.3in}\includegraphics[width=0.55\textwidth]{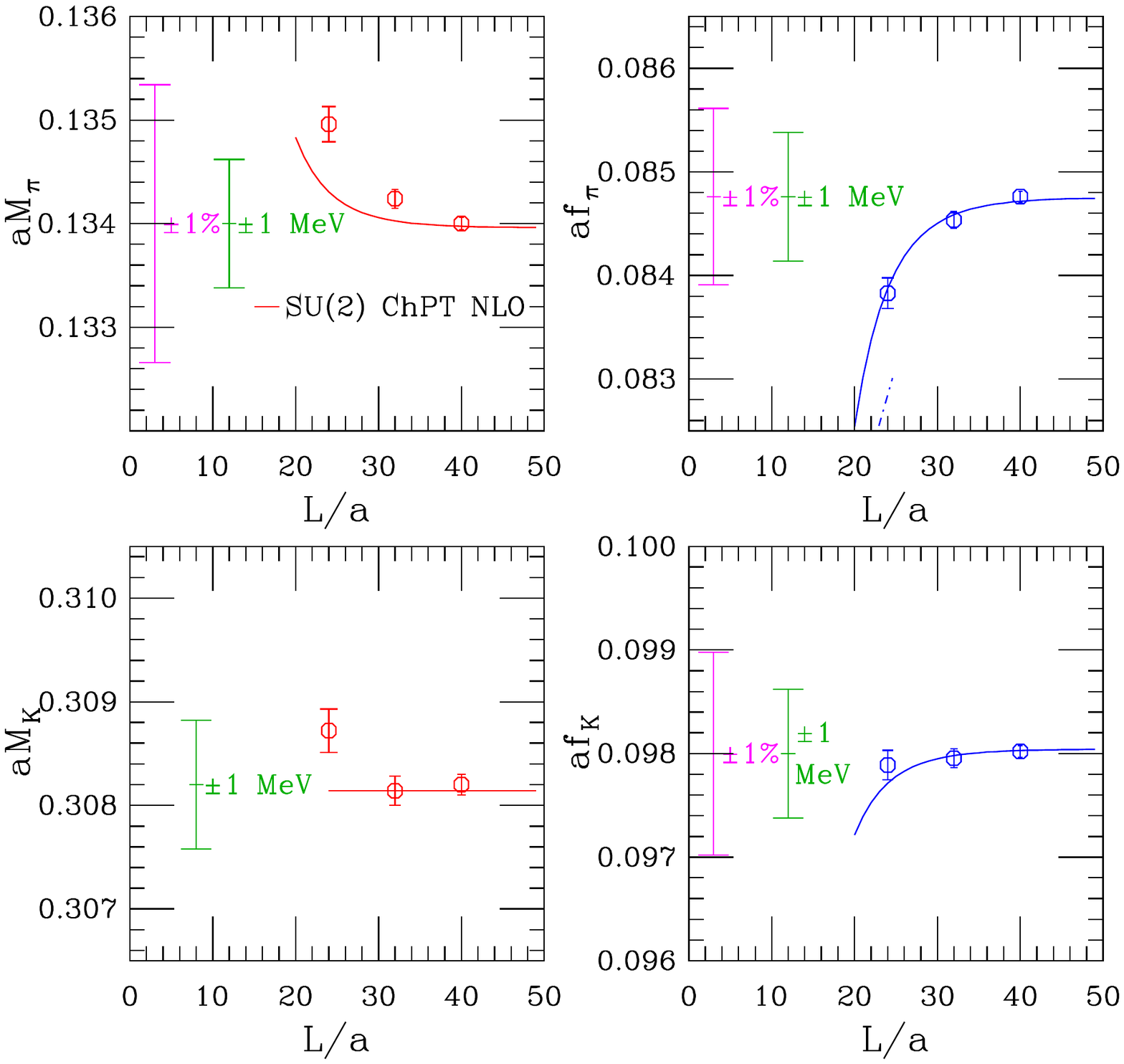} &
\hspace{-0.4in}\includegraphics[width=0.55\textwidth]{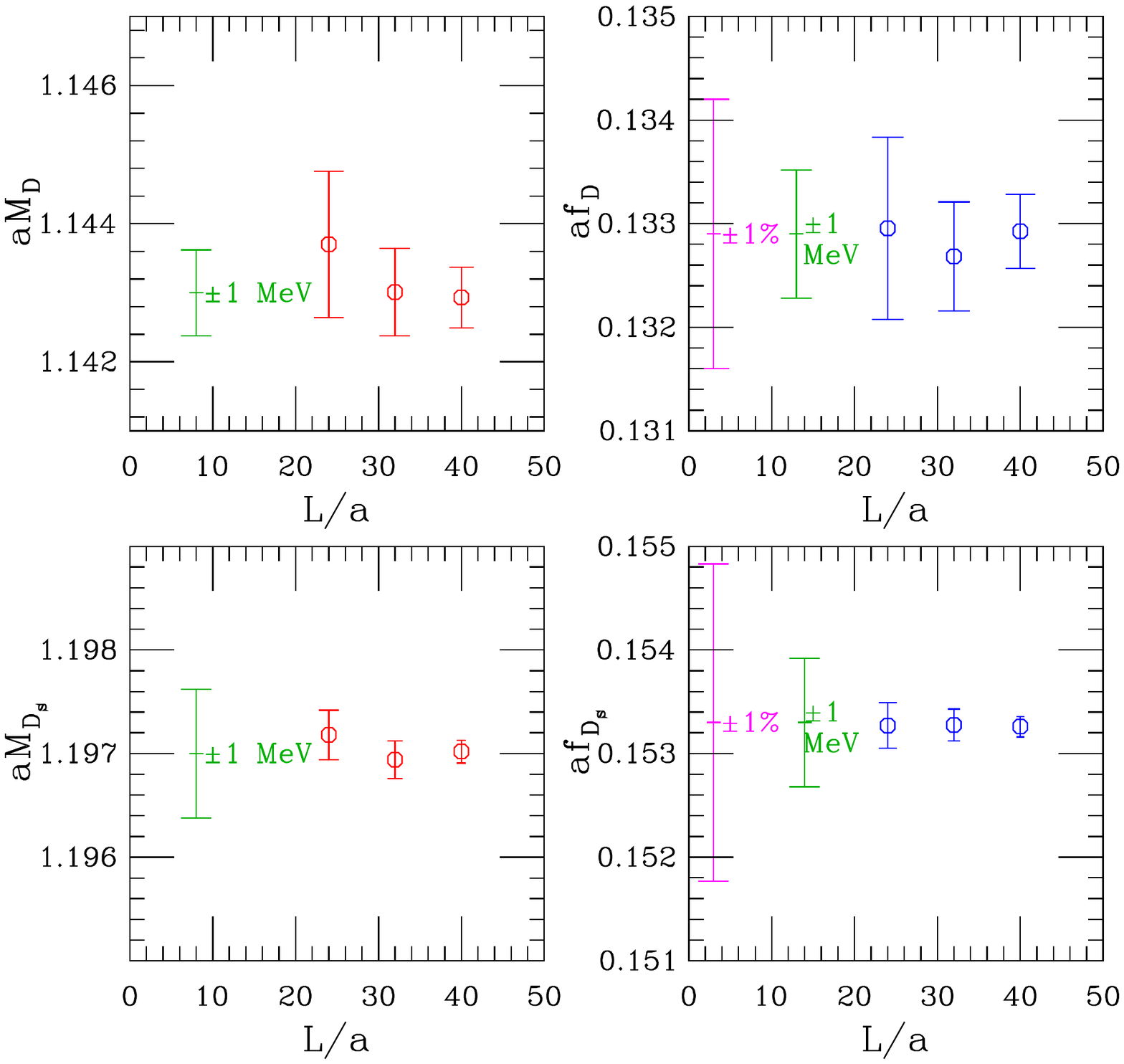} \\
% these plots were made by /home/doug/hisq/fpi/FD_FITTING/graphs/finitesize00507.csh
\end{tabular}
\end{center}
\vspace{-1.1in}
\caption{ \label{fig:finitesize}
Spatial size effects on $M_\pi$, $M_K$, $f_\pi$ and $f_K$ (left side),
and on $M_D$, $M_{D_s}$, $f_D$ and $f_{D_s}$ (right side).
Solid lines are the one loop chiral perturbation theory forms.
To show the magnitude of the effects, green error bars show 
an arbitrary value $\pm 1$ MeV, and magenta error bars $\pm 1$\%.
\vspace{-0.05in}
}
\end{figure}

Effects of finite spatial size were tested by running otherwise identical
ensembles with three different spatial sizes.  This was done at $a \approx 0.12$
fm with $m_l/m_s=1/10$.  The results,
summarized in Fig.~\ref{fig:finitesize}, show that the significant effects are
in the light pseudoscalar sector, on $M_\pi$, $f_\pi$ and $f_K$, and that
chiral perturbation theory is a reasonable guide.  From these results, we estimated
the effects on $M_\pi$, $f_\pi$ and $f_K$ at the physical light-quark mass
in a 5.5 fm box. (Since our fits are evaluated at this mass, it doesn't matter
that the ensembles with other sea-quark masses would have different finite-size
effects.) Then, in the analysis procedure described in section~\ref{STAGE2},
we used the values of $M_\pi$, $f_\pi$ and $f_K$ adjusted to this size box.
%(That is, we compute $\frac{f_K}{f_\pi(5.5\ {\mathrm{fm}})}$ {\it etc}.)
(That is, we compute ${f_K}/{f_\pi(5.5\ {\mathrm{fm}})}$ {\it etc}.)
Afterwards, we use these same factors to correct the computed $f_K$
%(or $f_\pi$, when we use $f_K$ to set the scale)
back to its infinite volume value.
We use the values with the finite-size adjustments
as our central values, and take one half of the shifts when these adjustments
are omitted as a remaining systematic error.

The effects of electromagnetic interactions and isospin breaking ($m_u \ne m_d$) are
considered together.
We tune using the $\pi^0$ mass, where electromagnetic and isospin breaking
effects are expected to be small.
Then we use a separate calculation of electromagnetic effects on the kaon
mass~\cite{EMREFERENCES}, whose main result is parameterized in terms of a violation
of Dashen's theorem  
\BNE M_{K^+,\mathrm{adj}}^2 = M_{K^+}^2 - \LP 1 + \Delta_{\mathrm{EM}} \RP \LP M_{\pi^+}^2 - M_{\pi^0}^2 \RP \ \ \ \ .\ENE
Here $M_{K^+,\mathrm{adj}}$ is the $K^+$ mass adjusted to remove the effects of electromagnetism,
and the result of Ref.~\cite{EMREFERENCES} is $\Delta_{\mathrm{EM}} = 0.65(26)$.
Then, in tuning the strange-quark mass from $2M_K^2-M_\pi^2$, we use the average squared
kaon mass $M_K^2 = \LP M_{K^+,\mathrm{adj}}^2 + M_{K^0}^2 \RP/2 $.
We also use this result combined with our meson mass measurements to determine the
up-down quark mass difference,
\BNE a^2 \LP M_{K^0}^2 - M_{K^+,\mathrm{adj}}^2 \RP = a \LP m_d-m_u \RP \PAR{(aM_K)^2}{am_l} \ \ \ \ , \ENE
where $\PAR{M_K^2}{m_l}$ is obtained from the difference in $M^2$ between the two lightest
valence-quark masses, with the strange mass set at the sea strange mass.
We estimate the remaining systematic errors from electromagnetism by combining,
in quadrature, the effects of changing $\Delta_{EM}$ by one standard deviation
with the effects of shifting the average kaon mass squared by 900 MeV$\null^2$, which
is the full result of a preliminary computation of electromagnetic effects
on this quantity~\cite{EMREFERENCES}.
When computing $f_D$, the light valence-quark mass is interpolated to the resulting down-quark
mass, and similarly for other quantities involving valence up and down quarks.
%A remaining systematic error from isospin breaking is taken to be one half
%of the shift in numbers when isospin violation is ignored.
We expect that the effects of electromagnetic interactions and isospin violation in the
sea quarks are small, and ignore them.

Errors from the scale setting and quark mass tuning procedure are estimated from
the change in results when $r_1$ or $f_{p4s}$ are used to set the scale instead of $f_\pi$.
%Errors from the scale setting and quark mass tuning procedure are estimated from
%using other procedures.  These include setting the scale using $f_K$ rather than $f_\pi$,
%using $r_1$ determined from the static quark potential (with a physical value determined
%from three flavor asqtad action simulations), and using an unphysical decay constant,
%$f_{p4s}$ from a meson with valence-quark masses of 0.4 times the strange-quark mass.
%When we use the alternative ``$f_K$'' scale setting scheme (or when we determine $f_K$ using the ``$f_pi$''
%scale setting), we make a similar adjustment of
%the input $f_K$ from its physical value, where the light quark is the up quark, to
%and adjusted value where the light-quark mass is $m_l = (m_u+m_d)/2$.

\begin{table}\begin{center}
\begin{tabular}{|l|l|l|l|l|l|}
\hline
Source         & $f_D$    & $f_{D_s}$ & $f_{D_s}/f_D$ & $m_c/m_s$ & $m_u/m_d$ \\
\hline
Statistics     & 3.0 MeV  & 0.5 MeV   & 0.016     & 0.041     & 0.009 \vspace{-3pt}\\
Excited        & 0.5 MeV  & 0.5 MeV   & 0.004     & 0.005     & 0.003 \vspace{-3pt}\\
Volume         & 0.3 MeV  & 0.2 MeV   & 0.0004    & 0.008     & 0.0005 \vspace{-3pt}\\ % 1/2 of shift
%Isospin        & 0.3 MeV  & 0.0 MeV   & 0.002     & 0.000     & na    \vspace{-3pt}\\ % 1/2 of shift, note sea isospin breaking too
E\&M           & 0.04 MeV & 0.12 MeV  & 0.001     & 0.001     & 0.021 \vspace{-3pt}\\  % shift lambda by .26, 901 MeV^2 for average m_K
Scale setting  & 2.0 MeV  & 1.3 MeV   & 0.003     & 0.02      & 0.023 \vspace{-3pt}\\ % diff. between Fpi and fp4s or r1
$a^2$ fit form & 2.9 MeV  & 3.3 MeV   & 0.010     & 0.09      & 0.011 \vspace{-3pt}\\ % larger diff between quad,a<=.15 and lin,a<=.12 fits or quad-phys
%Sys.          & 3.6 MeV  & 3.6 MeV   & 0.011     & 0.09      & 0.033 \\ combined systematic
\hline
\end{tabular}\\
\caption{ \label{tab:error_budget}
Error budget:
The statistical error includes error from valence-quark mass tuning and from
a continuum extrapolation with fixed fit form.
Estimation of the systematic error is discussed in the text.
%Excited state errors are from adjusting fit ranges and priors.
%Finite volume errors are one half of the shift from including the adjustments.
%Isospin breaking errors are from XXX.
%E\&M errors are from XXX.
%Scale setting errors are the full difference between the $f_\pi$ and $f_K$ scale settings.
%$a^2$ fit form errors are the larger of the full difference between the quadratic and linear fits or
%the difference between the quadratic and physical mass only fits.
% And my notes are in klingon:~/hisq/pubs/fd2012/re_systematics_oct26.txt
\vspace{-0.2in}
}
\end{center}\end{table}

\section{Conclusions} \vspace{-2.0mm}
\label{CONCLUSIONS}

%These results for the charmed meson decay constants $f_D$ and $f_{D_s}$ include
%the effects of four flavors of dynamical quarks, with larger spatial volumes
%than used previously.   The fermion action significantly reduces taste
%violating lattice artifacts relative to the asqtad action.  Systematic errors
%from addition excited states in the meson correlators, the choice of
%the fit for the continuum extrapolation, and uncertainties in the handling
%of isospin violation, electromagnetic effects and finite volume effects are
%included in our results.

We find preliminary results $f_D = 209.2(3.0)(3.6)$ MeV, $f_{D_s} = 246.4(0.5)(3.6)$ MeV,
$f_{D_s}/f_D = 1.175(16)(11)$, $m_c/m_s = 11.63(4)(9)$ and $m_u/m_d = 0.505(9)(33)$.
%Our results are consistent with those of other collaborations~\cite{FNAL95,HPQCD08,HPQCD10,ETMC09,
%PACS-CS11,ETMC11,FNAL11,ETMC12,HPQCD12}, and
Our results are consistent with those of other collaborations~\cite{ALLFD}, and
the errors are comparable to the most precise published calculation by HPQCD.
In the future we will further reduce the error by using
staggered chiral perturbation theory, which will allow us to make better
use of the correlators with unphysical valence-quark masses~\cite{KOMIJANITALK}, and by the
addition of an ensemble with physical sea-quark masses at a lattice
spacing of 0.06 fm.

%\begin{table}\begin{center}
%\caption{Preliminary results}
%\begin{tabular}{ll}
%$f_D = 209.2(3.0)(3.6)$ MeV & $f_{D_s} = 246.4(0.5)(3.6)$ MeV \\
%$f_{D_s}/f_D = 1.175(16)(11)$ \\
%$m_c/m_s = 11.63(4)(9)$ & $m_u/m_d = 0.505(9)(33)$ \\
%\end{tabular}\\
%\end{center}\end{table}

\acknowledgments \vspace{-2.0mm}

This work was supported by the U.S. Department of Energy and National
Science Foundation.
Computation for this work was done at
the Texas Advanced Computing Center (TACC),
the National Center for Supercomputing Resources (NCSA),
the National Institute for Computational Sciences (NICS),
the National Center for Atmospheric Research (UCAR), % Frost
the USQCD facilities at Fermilab,
and the National Energy Resources Supercomputing Center (NERSC),
under grants from the NSF and DOE.
%We thank WHO.

\end{document}